# Pupil Phase Series: A Fast, Accurate, and Energy-Conserving Model for Forward and Inverse Light Scattering in Thick Biological Samples

Herve Hugonnet[1,2,4], Chulmin Oh[1,2], Juyeon Park[1,2] and YongKeun Park[1,2,3,*]

[1]Department of Physics, Korea Advanced Institute of Science and Technology (KAIST), Daejeon 34141, South Korea
[2] KAIST Institute for Health Science and Technology, KAIST, Daejeon 34141, South Korea
[3] Tomocube Inc., Daejeon 34109, South Korea
[4] grougrou@kaist.ac.kr
* Y.K.P (yk.park@kaist.ac.kr)

We present the pupil phase series (PPS), a fast and accurate forward scattering algorithm for simulating and inverting multiple light scattering in large biological samples. PPS achieves high-angle scattering accuracy and energy conservation simultaneously by introducing a spatially varying phase modulation in the pupil plane. By expanding the scattering term into a Taylor series, PPS achieves high precision while maintaining computational efficiency. We integrate PPS into a quasi-Newton inverse solver to reconstruct the three-dimensional refractive index of a 180 µm-thick human organoid. Compared to linear reconstruction, our method recovers subcellular features—such as nuclei and vesicular structures—deep within the sample volume. PPS offers a scalable and interpretable alternative to conventional solvers, paving the way for high-throughput, label-free imaging of optically thick biological tissues.

## 1. Introduction

Accurate modeling of light propagation in complex media is essential for fundamental physics, advancing three-dimensional (3D) imaging, metrology, and computational microscopy [1-4]. In particular, reconstructing the refractive index (RI) distribution of complex objects such as thick biological samples remains challenging due to the presence of multiple scattering [5-7]. As light undergoes multiple forward scattering events, the assumptions underlying linear approximations—such as the Born or Rytov models—break down, leading to loss of resolution, contrast, and structural fidelity.

To address this, fast wave solvers such as the beam propagation method (BPM) [8,9] and multi-layer Born (MLB) [10] models have been widely adopted in imaging applications. BPM approximates the sample as a series of phase-modulating layers and propagates light under paraxial assumptions. MLB improves angle-resolved accuracy by applying first-order Born scattering within each slice. However, BPM is inaccurate for high-angle scattering and fails to model spherical aberrations, while MLB suffers from energy non-conservation and is sensitive to step size and RI contrast. While more accurate methods such as the convergent Born series (CBS) can rigorously model multiple scattering beyond the weak-scattering regime [11-13], they require repeated 3D convolutions and full-volume Green's function computations, making them computationally intensive.

In this work, we introduce the pupil phase series (PPS), a fast and accurate forward scattering model that overcomes the limitations of BPM and MLB. PPS models scattering as a spatially varying phase shift in the pupil plane, capturing high-angle diffraction while conserving energy. By expanding the scattering term as a Taylor series, the PPS method maintains computational efficiency comparable to multi-slice models while achieving accuracy similar to full-wave solvers.

We benchmark PPS against BPM, MLB, and CBS solver, using synthetic samples, demonstrating improved angular resolution and energy conservation. Furthermore, we integrate PPS into a quasi-Newton inversion framework to reconstruct the 3D RI distribution of a 180 μm-diameter human organoid. Our method reveals subcellular features obscured by multiple scattering in linear reconstructions, including deep nuclei and vesicular structures.

The PPS algorithm provides a generalizable and scalable forward model for multiple scattering, suitable for label-free imaging of optically thick biological samples. This paper is organized as follows: we first present the PPS formulation and its numerical implementation. We then evaluate its performance in simulation and demonstrate its application to experimental tomographic imaging of a large organoid. Finally, we discuss the broader impact of PPS across various imaging modalities.

## 2. Methods

### A. Forward Scattering using the Pupil Phase Series

Multi-slice methods simulate wave propagation through a volumetric sample by alternating between scattering and propagation steps. Let $n(x, y, z)$ be the 3D RI distribution, and $n_m$ the background RI. The incident field $E_0(x, y)$ is launched at $z = 0$, and the total field at depth $z_k = \delta z \cdot k$, $k \in [0, N_z -1]$ is denoted $E_k(x, y)$. The sample volume is discretized into $N_z$ axial slices of thickness $\delta z$, and lateral pixel size $\delta x = \delta y$.

The general multi-slice framework proceeds as follows:

**Table 1. Multi-slice algorithm**

```
for k = 1 to N_z − 1
    E_{z_k} = F^{-1} {U_{δz} F {E_{z_{k-1}}}}        → Propagation
    E_{z_k} = Scat(E_{z_k}, n_{z_k})                  → Scattering
end
```

Here, $U_{\delta z} = S \cdot e^{ik_z \delta z}$ is the refocusing kernel, $S = \delta(k_{xy} < 2\pi \text{NA}/\lambda)$ is the numerical aperture (NA) limit, $\delta(\cdot)$ the Dirac delta function, $\mathcal{F}$ the Fourier transform along the $x$ and $y$ axis, $k_{xy}$ the lateral spatial frequency in the Fourier space and $k_z = \sqrt{(n_m k_0)^2 - k_{xy}^2}$ the axial wave vector with $k_0 = 2\pi/\lambda$. Each multi-slice model defines the scattering functions $\text{Scat}(E_{z_k}, n_{z_k})$ differently:

*Beam Propagation Method (BPM)*

In the BPM, the field is updated by multiplying by the phase shift of the RI slice:

$$\text{Scat}_{\text{BPM}}\left(E_{z_k}, n_{z_k}\right) = E_{z_k} e^{ik_0 \left(n_{z_k}(x,y) - n_m\right)\delta z}.$$

This approximation is valid in the paraxial regime. For angled illumination a correction factor proportional to the inverse cosine of the illumination angle can be multiplied for improved accuracy. However, this does not improve the accuracy for scattered light propagating in directions different from the illumination. One of the advantages of the BPM method is that it ensures energy conservation of the transmitted field by modulating only its phase when the refractive indices are real.

*Multi-layer Born (MLB) method*

The MLB method uses a first-order Born approximation at each slice, where the scattered field is computed via a 2D convolution:

$$\text{Scat}_{\text{MLB}}\left(E_{z_k}, n_{z_k}\right) = E_{z_k} + \mathcal{F}^{-1}\left\{\frac{i}{2k_z} \mathcal{F}\left\{V\left(n_{z_k}\right) \delta z E_{z_k}\right\}\right\},$$

where $V(n_{z_k}) = k_0^2(n_{z_k}^2 - n_m^2)$ is the scattering potential. This approximation is valid for weak scattering $V(n_{z_k})\delta z \ll 1$ inside a given RI slice which imposes the uses of a small $\delta z$. Its advantage is that it can handle scattering at high angles beyond the paraxial regime. The ability to handle different scattering strength at different angles is granted by the product with the $1i/k_z$ term in the Fourier (pupil) space increasing scattering strength at high angles.

*Pupil Phase Series (PPS, proposed)*

The PPS model combines the angle-resolved accuracy of MLB and the energy conservation of BPM. We model scattering as a spatially varying phase modulation in the pupil (Fourier) plane. The PPS algorithm was devised by only changing the phase of the field but doing that at every spatial position at the pupil plane to handle high angle scattering:

$$\text{Scat}_{\text{PPS}}\left(E_{z_k}, n_{z_k}\right) = \iint_{x,y} E_{z_k}(x, y) P_{\text{PPS}}\left(n_{z_k}(x,y), x - x', y - y'\right) dx dy$$

$$P_{\text{PPS}}\left(n_{z_k}(x_2, y_2), k_{xy}\right) = \mathcal{F}\left\{P_{\text{PPS}}\left(n_{z_k}(x_2, y_2), x, y\right)\right\} = e^{ik_0 \left(n_{z_k}(x_2,y_2) - n_m\right)\delta z \frac{n_m k_0}{k_z}}$$

Due to the algorithm being a spatially varying convolution efficient implementation can be hard. Here we make use of a Taylor expansion on $P_{\text{PPS}}(m, k_{xy})$ to compute the scattering step as a sum of convolutions.

$$P_{\text{PPS}}\left(m, k_{xy}\right) = \sum_{0 \leq q} \frac{\left(ik_0(m - n_m)\delta z \frac{n_m k_0}{k_z}\right)^q}{q!}$$

$$\text{Scat}_{\text{PPS}}\left(E_{z_k}, n_{z_k}\right) = \sum_{0 \leq q} \frac{1}{q!} \mathcal{F}^{-1}\left\{\left(\frac{n_m k_0}{k_z}\right)^q \mathcal{F}\left\{E_{z_k}\left(ik_0\left(n_{z_k} - n_m\right)\delta z\right)^q\right\}\right\}$$

In this study, we use the extension to the third order $q \leq 3$. We note that this expression is equivalent to the BPM for $\text{NA} \ll 1$ and equivalent to the MLB for $n - n_m \ll 1$ it can handle high diffraction angles and conserves energy even for larger refractive indices.

**B. Inverse scattering using PPS**

To reconstruct the 3D RI distribution from measured multiple 2D field images, we integrate the PPS model into a quasi-Newton inverse scattering algorithm. Unlike gradient descent methods based on error backpropagation, quasi-Newton schemes linearize the forward problem at each iteration and directly solve the inverse problem—significantly accelerating convergence.

We denote $RI_k$ the RI at the $k^{th}$ step of the algorithm. PPS can be used to simulate the illumination and transmitted field propagation through $RI_k$. $E_{ill,3D} = PPS^+(E_{ill,m}, RI_k)$ denotes the simulation using the PPS algorithm to compute the volumetric field distribution $E_{ill,3D}$ for the $m^{th}$ illumination field $E_{ill,m}$ inside the $RI_k$. We also denote the measured transmitted field $F_{exp,m}$ and the scattering simulation in the backward direction $E_{trans,3D} = PPS^-(E_{trans,m}, RI_k^*)$. $PPS^-$ is similar to $PPS^+$ but propagation is carried in the opposite direction $\delta z \rightarrow -\delta z$ and "$k = 1$ to $N_z - 1$" $\rightarrow$ "$k = N_z - 1$ to $1$" in Table 1.

We further make use of a method similar to recently developed linear HT solver based on summation of the phase gradient [14,15] to perform the quasi-Newton update. We multiply the gradient by the illumination and transmitted intensity to avoid phase noise in the dark regions of the image. The gradients are finally integrated with spiral phase integration and deconvolved with a point spread function to update the RI. The algorithm is presented in Table 2.

**Table 2. Inverse scattering algorithm**

**for** $k = 1$ to outer iteration number
  $G = 0$
  **for** $m = 1$ to illumination number
    $E_{ill,3D} = PPS^+\left(E_{ill,m}, RI_k\right)$    $\rightarrow$ PPS simulations
    $E_{trans,3D} = PPS^-\left(E_{trans,m}, RI_k^*\right)$
    $W = C\left(E_{ill,3D}^* E_{trans,3D}\right)$
    $G = G + Im\left[W^* \mathcal{F}^{-1}\{k_x \mathcal{F}\{W\}\}\right]$    $\rightarrow$ phase gradient x
    $G = G + i\,Im\left[W^* \mathcal{F}^{-1}\{k_y \mathcal{F}\{W\}\}\right]$    $\rightarrow$ phase gradient y
  **end**
  $RI_{k+1} = RI_k + Re\left[\mathcal{F}^{-1}\left\{\dfrac{D}{k_x + ik_y} \mathcal{F}\{G\}\right\}\right]$    $\rightarrow$ spiral integration and deconvolution
**end**

Here $C(x) = A \cdot (\delta(|x| > 1)/|x| + \delta(|x| \leq 1)/\sqrt{|x|})$ is a normalization function. $D = OTF^* / (OTF \cdot OTF^* + \varepsilon)$ is the deconvolution operator and OTF is the vertical maximum of the 3D optical transfer function computed as the autocorrelation of the pupil function with $\varepsilon$ a regularization constant. For details, our implementation of the inverse scattering algorithm is available online[16].

## 3. Results and Discussion

### A. Simulation Benchmarking with a Phase Target

To benchmark PPS against existing multi-slice methods, we simulated wave propagation through a defined phase target (Fig. 1). The simulation target is composed of a thin scatterer (3 μm thickness, RI = 1.376) immersed in water (RI = 1.336) and placed behind a thicker cuboid scatterer (30 μm × 20 μm × 18 μm, RI = 1.376). This scenario was discretized onto a computational grid ($\delta x = \delta y = 0.12$ μm, $\delta z = 0.35$ μm; grid size: 400 × 500 × 172), illuminated by a plane wave parallel to the $z$-axis (Fig. 1a). The RI values range mimic realistic biological tissue properties.

The transmitted fields computed by PPS were compared against BPM, MLB, and an exact scalar reference solver (CBS). As shown in Figs. 1b–c, the PPS model exhibited significantly improved accuracy, closely matching the CBS results. Conversely, MLB exhibited poor energy conservation with energy increasing 70% (as computed from the mean intensity), leading to increased transmitted intensity through the cuboid. BPM maintained good energy conservation but failed to capture high-angle scattering accurately, resulting in noticeable errors at scatterer boundaries (Fig. 1d).

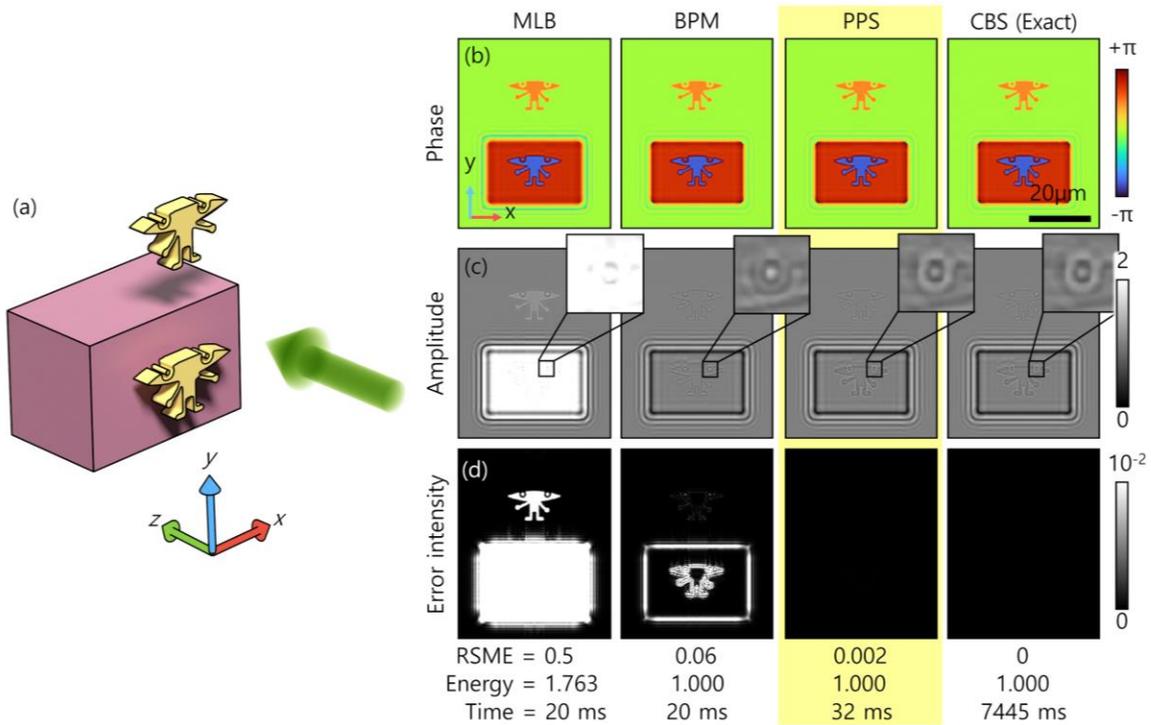

**Fig. 1. Scattering simulation benchmarking**. a) 3D scattering simulation layout. The green arrow shows the direction of illumination. Cuboid scatterer (pink color) and phase target (yellow color). b) Transmitted phase and b) amplitude for different simulation methods. d) Error intensity image in comparison with the CBS method e) Benchmarking of root mean square error and execution time.

Quantitatively, PPS achieved the lowest root mean square error (RMSE) among tested methods, indicating superior fidelity to CBS (Fig. 1e) while also achieving energy conservation up to the third digit. Notably, PPS maintained computational efficiency comparable to other multi-slice models, providing a 200-fold reduction in execution time compared to CBS. Thus, PPS offers a precise yet computationally efficient alternative suitable for large 3D imaging applications.

Compared to other high-order multi-slice models [17-21] that depend on computing axial derivatives, PPS offers simpler implementation. While our approach shares conceptual similarities with the wave propagation method [22,23], the series expansion introduced here significantly accelerates computational performance.

### B. 3D imaging of cell aggregates

We next assessed PPS's performance for inverse scattering problems by reconstructing 3D macrophage aggregates using holotomography (HT) combined with the PPS algorithm. Holotomography is a 3D quantitative phase imaging technique that visualizes the RI distribution in biological samples by inversely solving wave propagation. Although conventional HT methods often rely on linear scattering approximations, recent demands for imaging larger biological specimens such organoid [24], embryo [25] or thick histology tissues [26] have led to increased development of non-linear reconstruction approaches.

Typical non-linear reconstructions utilize error backpropagation [27] to compute gradients, followed by accelerated gradient descent algorithms [8,11,18]. Despite their accuracy, these methods often require numerous iterations for convergence. Alternatively, quasi-Newton algorithms [28,29] linearize and solve the imaging problem at each iteration, enabling faster convergence and better suitability for large-scale samples. This quasi-Newton approach underpins iterative optical diffraction tomography [30] and in-silico clearing quantitative phase gradient imaging [14,31] techniques, which are efficient and less computationally demanding compared to exact error backpropagation.

THP-1 cells (Korean Cell Line Bank) were cultured as per the supplier's protocol and used within 25 passages [32]. Cells were seeded onto collagen-coated substrates at a density of $10^5$ cells/mm² in complete RPMI 1640 medium (supplemented with 10% fetal bovine serum and 1% penicillin/streptomycin), differentiated into macrophages using 100 ng/mL phorbol 12-myristate 13-acetate (PMA) as previously described [21], and rested for 24 hours after replacing the medium [33].

The HT imaging system employed a Mach–Zehnder interferometer with a digital micromirror device (DMD) [34,35]. A coherent laser beam (457 nm wavelength, Cobolt Twist, Cobolt, Sweden) was split into reference and sample beams; the sample beam illuminated macrophage aggregates through a long working-distance condenser lens (LUMFLN60XW, Olympus, NA = 1.1). Scattered light was collected by an objective lens (UMFLN60XW, Olympus, NA = 1.1) and interfered with the reference beam to generate holograms captured by a CMOS camera (LT425M-WOCG, Lumenera, Canada).

Conventional linear reconstruction, limited by single-scattering approximations, generated blurred outlines and indistinct internal features, especially centrally located structures (Fig. 2a). Axial resolution was notably compromised, highlighting linear reconstruction limitations for highly scattering biological samples.

In contrast, iterative PPS-based reconstruction significantly enhanced image clarity and resolution. The resulting RI maps exhibited clear cell boundaries, lipid droplets, and distinct plasma membrane features (Fig. 2b, inset). Enhanced axial slices demonstrated precise localization of intracellular structures, confirming PPS's ability to mitigate multiple scattering effectively.

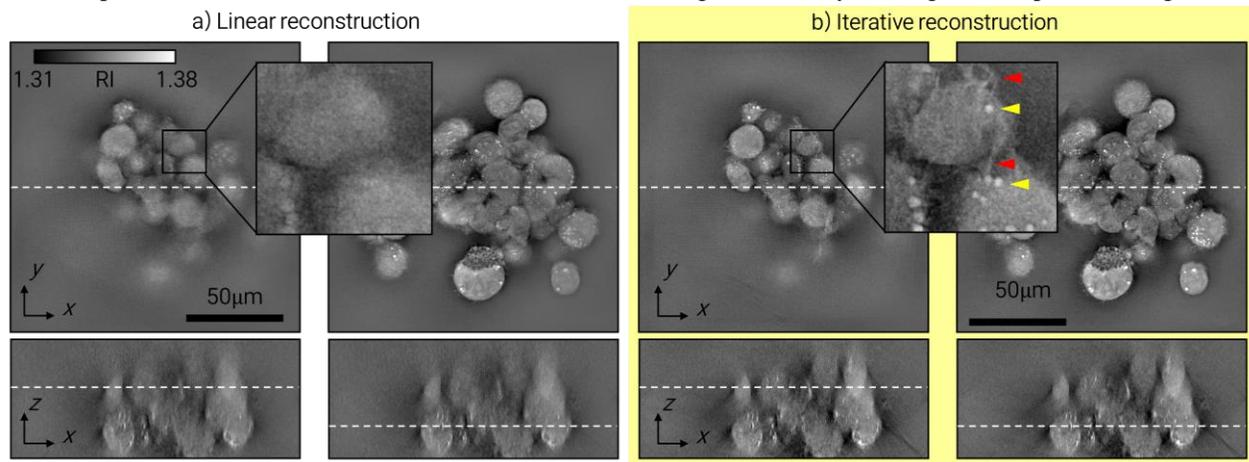

**Fig. 2. Comparison of linear and iterative tomographic reconstructions of 3D macrophage aggregates.** (a) Linear reconstruction shows blurred cellular boundaries and poor contrast within the interior of the macrophage cluster. Subcellular structures are indistinct, and the axial sections reveal low depth resolution. (b) Iterative reconstruction using the pupil phase series (PPS) algorithm significantly enhances structural clarity. The magnified inset highlights internal features such as plasma membrane (red arrowheads) and lipid droplet compartments (yellow arrowheads) that are not visible in the linear approach. Cross-sectional views in the *x–z* plane demonstrate improved depth localization and RI contrast.

## C. imaging of a large organoid

To further illustrate PPS's effectiveness in thicker samples, we imaged unlabeled live intestinal organoids using HT. Imaging was performed with a Mach–Zehnder interferometer setup. Illumination was provided by a 532 nm laser (Cobolt Samba), with angles controlled by a micro-electromechanical systems (MEMS) mirror (A5M24.3-2400AL-TINY48.4, Mirrorcle). Condenser (LUMFLN60XW, Olympus, NA=1.1) and objective lenses (UPLSAPO60XW, Olympus, NA=1.2) facilitated imaging onto a CMOS camera (CB654MG-GP-X8G3, Ximea, 300 fps, 1×4 decimation, 5024×5024 ROI). A total of 2000 projection angles were acquired, with iterative reconstructions progressively utilizing more projections (8, 16, 32, 64, 128, 256, 512, and finally 1024), compared to the linear method which used all 2000 projections.

PPS enabled efficient reconstruction of a large organoid volume (240 μm × 240 μm × 175 μm; 1995 × 1995 × 500 voxels) on a single GPU (RTX 3090, Nvidia). The 3dGRO human gastrointestinal organoids (Merck KGaA, cat# SCC335) were cultured in IntestiCult medium (STEMCELL Technologies), embedded in Matrigel domes within Tomodishes (Tomocube Inc.), fixed in 4% paraformaldehyde, and washed prior to imaging.

Both linear and iterative reconstructions effectively visualized the peripheral regions (Fig. 3 ii, vi); however, iterative reconstruction uniquely resolved deeper subcellular features, including clearly visible nuclei, spherical bubble-like lumen structures, and detailed organelles (Fig. 3 i, ii, iv, v).

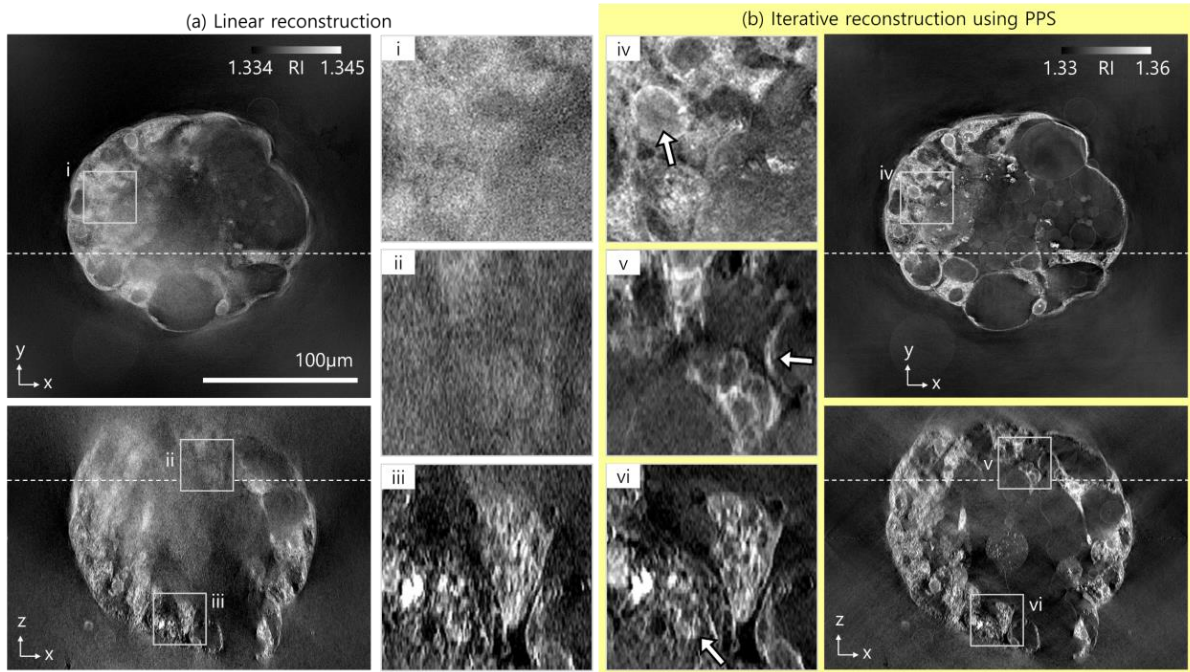

**Fig. 3. Tomographic reconstruction of an organoid.** a) In the linear reconstruction subcellular details are blurred deep into the sample. b) The iterative reconstruction recovers lost details. (i-vi) show enlarged regions of interest from a) and b) including a cell nucleus deep in the organoid (arrow iv), a spherical bubble-like structure at the top of the organoid lumen (arrow v) and subcellular organelles (arrow vi).

## 4. Conclusion

In this study, we introduced and rigorously demonstrated the PPS method, an advanced forward and inverse scattering model specifically designed to address challenges in imaging optically thick biological samples. PPS effectively combines the computational speed and simplicity of multi-slice models with the accuracy and energy conservation characteristics essential for handling multiple scattering events, particularly at high diffraction angles. By implementing a spatially varying pupil plane phase modulation and expanding the scattering term into a Taylor series, PPS achieves superior accuracy without significantly increasing computational complexity.

We validated PPS's performance through extensive numerical simulations, confirming its improved accuracy and computational efficiency relative to traditional methods such as the BPM and the MLB approach. PPS demonstrated excellent agreement with the exact scalar convergent Born series solver while maintaining execution speed hundreds of times faster, positioning PPS as a highly practical and efficient solution for large-scale 3D imaging applications.

The integration of PPS into a quasi-Newton inverse solver facilitated high-resolution three-dimensional reconstructions of biologically relevant samples, including macrophage aggregates and thick human gastrointestinal organoids. Notably, iterative PPS reconstructions successfully resolved detailed subcellular structures that were indistinguishable using linear reconstruction methods due to multiple scattering effects. These reconstructions demonstrated PPS's unique capability to visualize deep intracellular features, significantly enhancing both axial and lateral resolutions.

Given the increasing demand for accurate and efficient imaging of large, complex biological specimens, PPS represents a significant advancement in computational microscopy. Its scalability, computational efficiency, and straightforward implementation make PPS broadly applicable beyond refractive index tomography, potentially benefiting fields such as scattering modeling [36-38], fluorescence [39,40], electron and X-ray microscopy [41,42], ultrasound imaging [43], wavefront shaping [44], and even nanophotonics design [13,45].

## 5. List of abbreviations

Refractive index (RI)
Pupil Phase Series (PPS)
Beam Propagation Method (BPM)
Multi-layer Born (MLB)
Convergent Born series (CBS)
Holotomography (HT)

Root mean square error (RMSE)
Micro-electromechanical systems (MEMS)

**Back Matter**

**Funding.** This work was supported by Tomocube Inc., National Research Foundation of Korea grant funded by the Korea government (MSIT) (RS-2024-00442348, 2022M3H4A1A02074314), Korea Institute for Advancement of Technology (KIAT) through the International Cooperative R&D program (P0028463), and the Korean Fund for Regenerative Medicine (KFRM) grant funded by the Korea government (the Ministry of Science and ICT and the Ministry of Health & Welfare) (21A0101L1-12).

**Authors' contributions** H.H. and C.O. developed the theory, H.H. analyzed the data, H.H. and J.P. conducted the experiments, and Y.P. supervised the project. All authors wrote the manuscript.

**Competing interests.** H.H., J.P., and Y.P. have financial interests in Tomocube Inc., a company that commercializes HT instruments and is one of the sponsors of the work.

**Availability of data and materials.** Data and code underlying the result in this paper are available online [16].